\documentclass[11pt,twoside]{article}

\usepackage{asp2006}
\usepackage{epsf}
\usepackage{lscape}

\begin{document}
\title{The evolving disk galaxy population}   
\author{Eric F.\ Bell}   
\affil{Max-Planck-Institut f\"ur Astronomie, K\"onigstuhl 17, D69117 Heidelberg, Germany}    

\begin{abstract} 
In this contribution, I present a simplified overview
of the evolution of the disk galaxy population since $z=1$, and a brief
discussion of a few open questions. 
Galaxy evolution surveys have 
found that the disk galaxy population forms stars intensely
at intermediate redshift.  In particular, they dominate the cosmic 
star formation rate at $z<1$ --- 
the factor of ten drop in cosmic average comoving star formation rate
in the last 8 Gyr is driven primarily by disk physics, not by a
decreasing major merger rate.  Despite this intense star formation, 
there has been little change in the stellar mass density 
in disk galaxies since $z=1$; large numbers of disk galaxies
are being transformed into non-star-forming spheroid-dominated
galaxies by galaxy interactions, AGN feedback, environmental effects, 
and other physical processes.  Finally, despite this 
intense activity, the scaling relations of disk galaxies 
appear to evolve little.  In particular, as individual galaxies
grow in mass through the formation of stars, they appear to grow
in radius (on average, the population grows inside-out), and they 
appear to evolve towards somewhat higher rotation 
velocity (i.e., mass is added at both small and large radii
during this inside-out growth).  
\end{abstract}



\section{Introduction}

The properties and evolution of the disk galaxy population are a 
critical diagnostic of the galaxy formation process in a dark 
matter-dominated Universe.  Galaxy disks, formed naturally as a result
of a collapse in which at least some of the angular momentum is 
conserved \citep{fall80}, are relatively fragile beasts.
Interactions with low-mass halos are likely to thicken 
the stellar disk \citep[e.g.,][]{toth92,benson04} and/or lead 
to warps and lopsidedness; interactions
with galaxies/halos with masses within a factor of a few of the galaxy 
mass look likely to disrupt the stellar disk entirely 
\citep[e.g.,][]{koda08}.  Thus, the evolution and properties
of the disk galaxy population are not only a fascinating puzzle in 
their own right, but also give insight into those environments 
which are the least affected by the characteristic pummeling that 
galaxies in $\Lambda$CDM seem to receive.

In the seven years since the first Vatican disk galaxies meeting, 
a number of the basic observational features of the evolution of the disk 
galaxy population have come into place.  In this contribution, 
I present a brief overview of what I consider to be some of the 
most important features of the evolution of the disk galaxy population 
since $z=1$ (roughly the last 7-8 billion years), and pose a few
questions that I had after this conference.  These features fall under
three broad themes: star formation in disks, stellar mass in disks, 
and disk galaxy scaling relations.  In what follows, in the 
spirit of a review, I will deliberately
over-simplify the observational results somewhat in order 
to clarify the three basic messages of this contribution.

\section{Star formation in disks}

As a community, we have made excellent progress towards
understanding the contribution of disk galaxies to the 
star formation census of the cosmos, especially 
since $z=1$.  While clearly much work remains to be done, 
I would maintain that the bottom line is clear: since 
$z=1$, the majority of all stars that form do so in 
galactic disks \citep[see, e.g.,][and Fig.\ \ref{fig}]{hammer05,bell05,wolf05,melbourne05}. 

While this bottom line, I think, is clear, there are 
a number of issues with current observational studies 
which need to be understood and improved upon.  
The key issue is that identifying disks 
unambiguously is not straightforward: 
does one do this using a cut in light profile concentration, 
Sersic Index (essentially selecting galaxies with light profiles
not that different from an exponential), using visually
classified disks, or by splitting by rest-frame color
into blue cloud and red sequence?  Workers in this field 
have taken a number of different approaches, and it is a
testament to the robustness of the result that independent of
approach it is clear that the bulk of star formation is in disks.  Yet, 
in order to understand this result in depth, more refined approaches
will be necessary.  Furthermore (as I will touch on later), there 
are observational selections (in photoz or spectroscopic z studies) that
favor the detection of low mass galaxies with preferentially 
high SFRs; such a selection will skew our understanding of the population 
average.  Finally, high-quality SFR estimates from far-IR SEDs
will come into place in the next years with the upcoming launch of 
{\it Herschel}.  

The following is an incomplete list of a few 
implications of this result.
\begin{itemize}
\item Disk galaxies are forming stars vigorously at 
intermediate redshift.  In the particular case of massive galaxies, 
star formation happens at rates typically high enough to 
allow them to double their mass in just a few Gyr \citep{noeske07a,noeske07b,zheng07}.
\item The fact that most of the star formation is in disks 
at $z<1$ makes it clear that most of the star formation 
is {\it not major merger-driven}.  Jason Melbourne and coworkers
presented a poster at the meeting in which they argued
for an even stronger statement --- they found in deep imaging 
of star-forming disks relatively little evidence for 
preferentially more minor merging/accretion in intensely 
star-forming systems relative to non-star-formers (although \citealp{hammer07} 
have a different view, based on disturbed velocity fields in 
a significant fraction of intermediate redshift LIRGs).
\item Thus, the drop in the cosmic 
SFR from $z=1$ to the present day is due primarily to 
`quiescent' processes, e.g., gas consumption (the consumption of 
gas that the galaxy already has) and/or a reduced rate of accretion 
of fresh gas from the cosmic web (either through cooling from a hot medium
or from direct accretion). 
\end{itemize}

Given these implications, there are a few questions that 
would be great to have understood in more detail:
\begin{itemize}
\item Estimates of star formation history in the solar cylinder
tend to come up with a roughly constant 
SFR since the formation of the disk.  Study of the 
evolution of the SFR of the disk galaxy population 
since $z=1$ suggest at least factors of several higher SFR
7-8\,Gyr ago, on a global scale.  It will be an important
challenge for the community to `close the loop', and to 
verify (or rule out!) that the global SFR of the Milky Way was 
significantly higher in the past, and that one expects
to infer a $\sim$constant SFR at $\sim$3 disk scale lengths
(where dynamical re-arrangement of stars after they 
form may play an important role here; see, e.g., the contribution 
of Debattista and Ro\v{s}kar; \citealp{roskar07}).
\item From work with high velocity {\sc Hi} clouds, there is the suggestion 
of an infall rate of a few tenths of a solar mass per year of new gas
(Putman, this conference; \citealp{peek07}); 
this is less than half of the current SFR.  It will be an important 
challenge to better measure the infall rate today, and better constrain
the infall history of disk galaxies in a general sense.  This is a 
field in which there is excellent promise, as it is an ideal 
arena for the synthesis of observations of SFRs, metallicities (and metallicity histories), and gas contents (especially with ALMA) with simulations of 
disk galaxy evolution in a cosmological context.
\item It would appear that the {\it average} specific SFR of low-mass galaxies
($\le 10^{10}M_{\odot}$ in stars) at intermediate redshifts is
very high (doubling times are typically 1\,Gyr; s
ee Fig.\ 2 of \citealp{zheng07}).
I am sure that this is naive observer-speak, but I am always 
troubled by an {\it entire population}, and not a small one, 
which appears to be 
star forming itself out of existence.  There are simply not enough high
mass galaxies around to represent the end-products of 
this intense star formation.  I suspect that our census of 
low mass objects at intermediate redshifts is very biased towards only those 
galaxies with high SFRs.  If incompleteness really is the problem, 
rectifying this situation 
will not be straightforward, and will require a dedicated 
near-IR photoz effort (to reduce the effects of
star formation on the observed luminosities) 
followed by deep and determined spectroscopy (to verify the 
often unreliable photozs of very faint objects).  
\end{itemize}

\section{Stellar mass in disks}

\begin{figure}[t]
\plotone{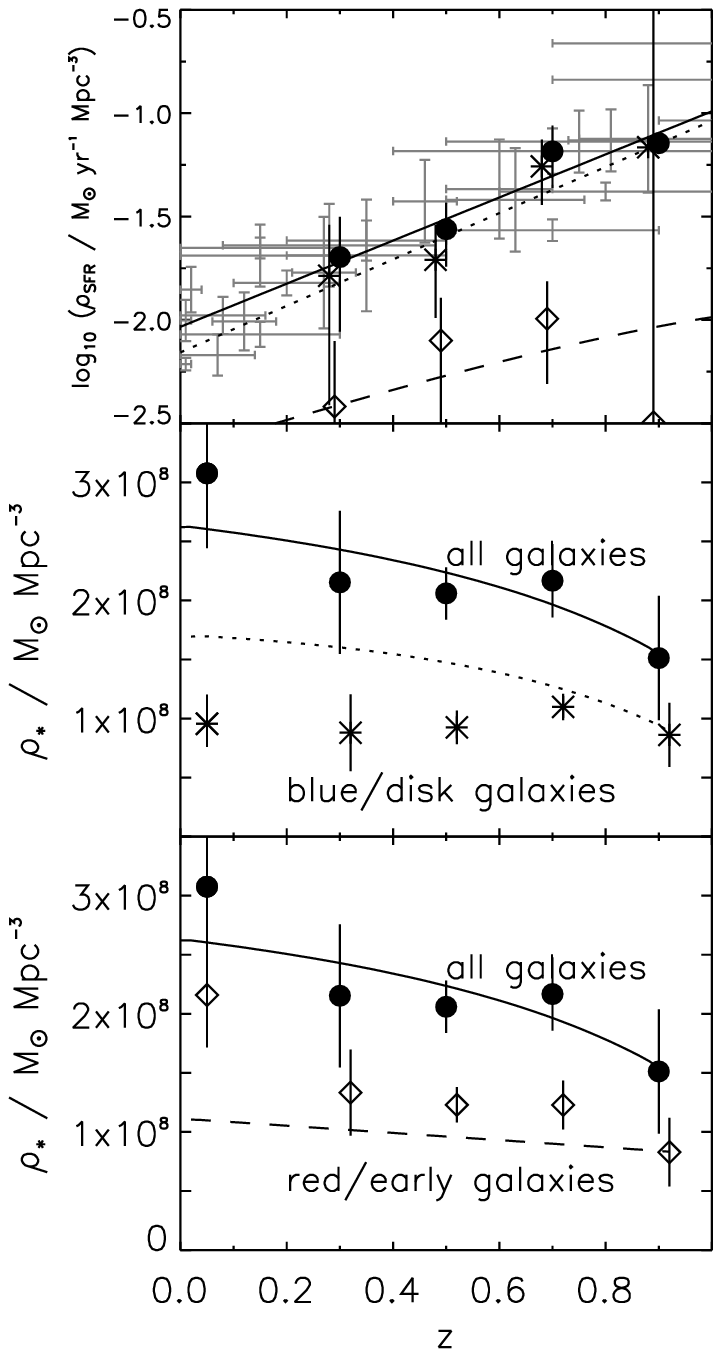}
\caption{{\bf Top:} The cosmic SFR history since $z=1$ for 
all galaxies (filled circles and a compilation of literature values
in gray).  The contribution from blue cloud (primarily disk 
galaxies) is shown as asterisks. 
The red sequence contribution (mostly early-type galaxies
with no star formation -- the star formation is from a mix of mergers
and edge-on spirals) is shown as diamonds.
{\bf Middle:} The filled circles show the stellar mass evolution 
predicted from the cosmic SFR (solid line) along with observations
from \citet{borch}.  Since almost all star formation is in blue 
(primarily disk) galaxies, the predicted growth in stellar
mass in the blue galaxy population is substantial (dotted line); 
there is no observational evidence for such a mass growth (asterisks).
{\bf Bottom:} Instead, all observed mass growth is in red
sequence (primarily early-type) galaxies (diamonds; almost
no growth is predicted as shown by the dashed curve).
Adapted from \citet{bell07}.} \label{fig}
\end{figure}

An obvious implication of the above discussion is that we should 
expect to see a huge increase in the stellar mass density 
in disks since $z=1$ --- they are forming stars like crazy; they have to 
go somewhere (see the predicted mass growth in blue galaxies -- dotted line in the middle panel of in Fig.\ \ref{fig})!  

The last years have seen the creation of a number of 
datasets with which one can estimate the stellar mass function 
of disks (bearing in mind the obvious issue of `What is a disk?'
that was brought up in the last section).  The key result of 
these studies is that there has been {\it very little evolution}
in the mass density in disk-dominated (or blue cloud\footnote{A selection 
for blue cloud galaxies is often made in this game, because the surveys
which have enough volume to measure the relatively subtle changes
in the stellar mass density were not large enough to have 
HST imaging covering their entirety.  Thus, blue cloud is 
often selected, noting that almost all blue cloud 
galaxies are disk dominated (e.g., \citealp{bell04a}).}) galaxy 
population since $z=1$.  Instead, one sees the stellar mass
density of non-star-forming (primarily bulge-dominated) galaxies
growing by factors of two or more in that period of time
\citep[e.g.,][and Fig.\ \ref{fig}]{bell04b,brown06,faber07,bell07}\footnote{For those who are more empirically-minded, 
the observable result goes as follows.  The rest-frame 
$B$-band luminosity function evolves by roughly 1 magnitude
in $L^*$ (putting it in terms of a Schecter function fit to the luminosity
function) since $z=1$ for both red and blue galaxies.  The density
normalization $\phi^*$ remains basically unchanged for blue galaxies, 
whereas for red galaxies it appears to have increased by roughly a factor
of 2 or more; the end result is that the luminosity density in 
blue disks was 2.5 times higher at $z=1$ than today, and the luminosity
density in red  early-type galaxies remains basically unchanged from $z=1$ 
to the present.  In this time, both red and blue galaxies have 
reddened considerably (because the stars in these galaxies
have aged from $z=1$ to $z=0$).  It is unavoidable that 
as the population ages and becomes redder that it should fade
also.  The predicted fading (from stellar population models)
for both red and blue galaxies is 
roughly $\sim 1$\,mag since $z=1$ (corresponding more or less
to the observed shift in $L^*$; i.e., the knee in the stellar
mass function does not shift appreciably since $z=1$).  Thus, 
the stellar mass density in blue disks appears to remain more 
or less constant since $z=1$, whereas red early-types 
appear to double or more in mass density --- in both cases, 
this result is driven by the changes in $\phi^*$ rather than 
changes in the basically non-evolving 
knee of the stellar mass function $M^*$. }.  

The implication of this is clear, and profound.  There
are physical processes which are turning off 
star formation in a significant fraction of star-forming
disk galaxies, and leading to the creation or augmentation of 
a bulge.  This is not a subtle process, as it has affected 
$\sim$1/2 of the stellar mass density which either was on, or 
formed in, the blue cloud. 

There are a number of possible physical processes
that are likely to play a role: e.g., galaxy merging (to create bulges), 
AGN feedback (to evacuate the gas, or prevent it from 
cooling at late times), stripping of gas in group/cluster
environments, and tidal stripping/harassment in dense 
environments.  A key issue for the community
in the future is to identify and explore the relative
importances and roles of these processes on shaping the 
evolving disk galaxy population.

\section{The evolution of galaxy scaling relations}

OK, where are we?  We've seen that disk galaxies form stars
intensely since $z=1$.  We've seen that many disk galaxies
are transformed into non-star-forming spheroid-dominated galaxies
since $z=1$.  There has been intense activity in the disk 
galaxy population since $z=1$ --- it would be reasonable
to expect some signature of the physics of this intense
activity imprinted in the galaxy scaling relations.

While there is much work left to do, I would maintain 
that we have a basic understanding of the evolution of 
the stellar mass Tully-Fisher relation (the 
stellar mass--rotation velocity relation) and the 
stellar mass--radius relation.  It would appear 
that neither relation changes appreciably
since $z=1$ \citep[e.g.,][]{barden05,con05,trujillo06,flores06,kassin06}.  

These are key observational results, as they places significant
constraints on the evolution of galaxies:
\begin{itemize}
\item Disk galaxies appear to evolve {\it along} the stellar mass--radius
relation.  They are forming stars, therefore they are growing significantly 
in stellar mass since $z=1$.  Thus, the population must be growing in radius
as the population builds up mass (in order to stay on the scaling relation) 
--- the population is on average
growing inside-out \citep{barden05,trujillo06}.
\item Disk galaxies appear to also evolve {\it along} 
the stellar mass--rotation velocity relation.  Thus, as they grow
in stellar mass they also grow in rotation velocity.  This is an 
interesting constraint.  The measured rotation velocity is set by the dark
and luminous matter content within roughly 2 half-light radii.  
If galaxy mass is added at {\it only} large radii (i.e., really inside-out
growth), one would expect rotation velocities to change little 
as the galaxy gains mass.  Thus, the non-evolution of the 
stellar mass Tully-Fisher relation shows that galaxies are 
adding mass at all radii (affecting the rotation velocity), 
while making sure to preferentially add it at large radii
(to ensure inside-out growth). 
\end{itemize} 
Models of galaxy formation which are able to 
start interpreting such phenomenologies are 
emerging (e.g., Somerville, this conference; \citealp{somer08}), 
and will help to both sharpen our understanding 
of the evolution of the disk galaxy population, and motivate
improved analyses of new datasets.

\section{Conclusions and discussion}

In this contribution, I have presented an over-simplified
view of the evolution of the disk galaxy population:
\begin{itemize}
\item Disk galaxies form stars intensely since $z=1$.  They dominate
the cosmic average SFR since $z=1$; the factor of 10 decline in 
cosmic SFR since $z=1$ is primarily driven by disk galaxy physics (gas 
consumption, decreased accretion from the cosmic web), not 
changes in the galaxy merging rate.
\item Despite this intense star formation, the disk galaxy population 
does not grow in mass since $z=1$.  Large numbers of star-forming
disks are being transformed into non-star-forming spheroid-dominated
galaxies.  
\item As disk galaxies gain mass through star formation, 
they grow in radius (i.e., the population grows inside-out) 
and rotation velocity (i.e., mass is added at small and large 
radii during this inside-out growth).  
\end{itemize}
A final word.  This picture of a dynamic disk galaxy population 
has important implications for how we interpret the results of our
observations.  Statements such as 'massive disk galaxies
are in place at $z=1$ and appear to evolve little to the present 
day' are incorrect.  Some of these massive disk galaxies have become
bulge-dominated at the present day (and are therefore not 
in the $z=0$ control disk galaxy sample), some are in 
(rare) even more massive disk galaxies.  This has implications
for how one should interpret a variety of observations: 
bar fractions, metallicity--mass relations, mass--radius and 
mass--rotation velocity relations.  While some studies
specifically address such issues (e.g., \citealp{kassin06}), it is important
to bear the (observed!) dramatic evolution of the disk galaxy population  
in mind when interpreting scaling relations and properties of the 
evolving disk galaxy population.

\acknowledgements 

I wish to thank the organizers of this conference for their 
invitation to attend this conference, and the conference participants
for creating (what at least I felt to be) a lively atmosphere.
I wish to thank Mary Putman, Christine Chiappini, 
Luc Simard, Ken Freeman, John Kormendy, Roelof de Jong, and 
St\'ephane Courteau for thought-provoking conversations at 
the conference, some of which have shaped this conference proceedings.
I wish to thank my closest collaborators on disk 
galaxy work in recent years --- Marco Barden, Dan McIntosh, Christian Wolf, 
Roelof de Jong, and Hans-Walter Rix.  I wish to 
thank also the DFG (German Science Foundation) for their support
in the form of a grant from the Emmy Noether Programme.  Finally, 
I wish to thank Starbucks, Heidelberg for their hospitality while these
conference proceedings were being written.


\end{document}